\documentclass[reprint,showpacs,amssymb,amsmath,aps,pra]{revtex4-1}
\usepackage{graphics}
\usepackage{graphicx}
\usepackage{amssymb}
\usepackage{amsmath}
\usepackage{bm}
\begin{document}
\title{Quantum discord and information deficit in spin chains}
\author{N. Canosa, L. Ciliberti,  and R. Rossignoli}

\affiliation{
Departamento de F\'{\i}sica-IFLP-CONICET-CIC, Universidad Nacional de La Plata, \\
C.C.67, La Plata (1900), Argentina}

\begin{abstract}We examine the behavior of quantum correlations of spin pairs in a
finite  anisotropic $XY$ spin chain immersed in a transverse magnetic field,
through the analysis of the quantum discord and the conventional and quadratic
one way-information deficits. We first provide a brief review of these
measures, showing that the last ones can be obtained as particular cases of a
generalized information deficit based on general entropic forms. All these
measures coincide with an entanglement entropy in the case of pure states, but
can be non-zero in separable mixed states, vanishing just for classically
correlated states. It is then shown that their behavior in the exact  ground
state of the chain exhibits similar features, deviating significantly from that
of the pair entanglement below the critical field. In contrast with
entanglement, they reach full range in this region, becoming independent of the
pair separation and coupling range in the immediate vicinity of the factorizing
field. It is also shown, however, that significant differences between the
quantum discord and the information deficits  arise in the local minimizing
measurement that defines them. Both analytical and numerical results are
provided.
\end{abstract}
\maketitle

\section{Introduction}

The investigation of non-classical correlations in mixed states of composite
quantum systems has attracted strong attention in recent years.  While in pure
states such correlations can be identified with entanglement
\cite{S.35,Sch.95,NC.00,Ve.06,HR.07}, in the case of mixed states, separable
(unentangled) states, defined in general as convex mixtures of product states
\cite{RF.89}, i.e., as states which can be generated by local operations and
classical communication (LOCC), may still exhibit non-classical features. The
latter emerge from the possible non-commutativity of the different products and
lead, for instance, to a finite value of the quantum discord
\cite{Zu.01,HV.01,Zu.03} and other recently introduced related quantifiers of
quantum correlations \cite{Mo.10,Mo.11}. These quantifiers  include the one-way
information deficit \cite{HHH.05,SKB.11,Zu.03}, the geometric discord
\cite{DVB.10}, generalized entropic measures \cite{RCC.10,RCC.11} and more
recently the local quantum uncertainty \cite{LQU.13,LU.12} and the trace
distance discord \cite{TDD.13,Hu,Na,Ci}. While entanglement is certainly
necessary for quantum teleportation \cite{tel} and for an exponential speed-up
in pure state based quantum computation \cite{JL.03}, interest on these new
measures has been triggered by the existence of mixed state based quantum
algorithms like that of \cite{KL.98}, able to achieve an exponential speedup
over the best classical algorithms for a certain task, with vanishing
entanglement \cite{DFC.05} but finite quantum discord \cite{Ca.08}. And various
operational interpretations of the quantum discord and other related measures
have been provided \cite{AD.12,KW,MD.11,CAB.11,Fan,Mo.11,LQU.13,Na,Aex}.

In this article we will concentrate on the quantum discord
\cite{Zu.01,HV.01,Zu.03} and the generalized entropic measures of
\cite{RCC.10}, which include as particular cases the von Neumann based one-way
information deficit \cite{HHH.05,SKB.11,Zu.03} and the geometric discord
\cite{DVB.10}, and which represent a generalized information deficit. The
quantum discord as well as all other related measures require a rather complex
minimization over a local measurement or operation  which has limited their
applicability to small systems or special states. The optimization problem for
the quantum discord was in fact recently shown to be NP complete \cite{NP}. The
advantage of the generalized entropic formalism is, first,  the possibility of
using simpler entropic forms like the linear entropy, which, as will be
discussed in section 2, enables an easier evaluation (it does not require the
diagonalization of the density matrix) and a more direct experimental access
(it can be determined without a full state tomography). This entails that an
explicit solution of the associated optimization problem for certain states can
be achieved. The generalized formalism also allows to identify some universal
properties, i.e.\ valid for any entropic form (and not just for a particular
choice of entropy) satisfied by the post-measurement state.

We first provide in section 2 an overview of the main concepts and properties
associated with these measures. We then apply these measures to examine the
quantum correlations of spin pairs in the exact ground state of finite spin
$1/2$ chains with $XY$-type couplings in a transverse magnetic field, through
their entanglement, quantum discord and information deficit. All separations
between the pairs are considered. Several important studies of the quantum
discord in  spins chains have been made
\cite{spin.08,Sa.10,CRC.10,We.11,LSZ,RCM.12,CA.13,CCR.13,Huang.14},  but the
relation with the generalized information deficit and the differences between
their optimizing measurements in these spin pairs have not yet been  analyzed
in  detail. We have recently investigated these aspects for an $XX$ spin chain
in \cite{CCR.13}, and will here extend this analysis to the  anisotropic $XY$
case. It is  first shown that in contrast with the pair entanglement, the
quantum discord and the information deficit exhibit, for the exact ground state
of these chains, common features such as an  appreciable finite value below the
critical field, for all separations. Moreover, they  approach a finite {\it
common non-zero value} \cite{CRC.10} at the remarkable factorizing field
\cite{Kur.82,Am.06,RCM.08,RCM.09,GAI.09,CA.13} that these  chains can exhibit
in the anisotropic case. On the other hand, we will also show that important
differences between the quantum discord on the one side, and the standard and
generalized information deficit on the other side, do arise in the minimizing
local spin measurement that defines them. While in the quantum discord the
direction of the latter is always orthogonal to the transverse field, in the
other measures it exhibits a perpendicular to parallel transition as the field
increases, which is present for all separations  and which reflects significant
qualitative changes in the reduced state of the pair. This difference indicates
a distinct response of the minimizing measurement of these quantities to the
onset of quantum correlations.

\section{Measures of quantum correlations}
\subsection{Quantum Entanglement}

We start by providing a brief overview of the basic notions. A pure state
$|\Psi_{AB}\rangle$ of a bipartite system $A+B$ is separable iff (if and only
if) it is a product state $|\Psi_A\rangle|\Psi_B\rangle$. Otherwise it is
entangled. The Schmidt decomposition \cite{NC.00}
\begin{equation}|\Psi_{AB}\rangle=\sum_{k=1}^{n_s}\sqrt{p_k}\,|k_A\rangle|k_B\rangle\,,
\label{1SD}\end{equation}
where $|k_{A(B)}\rangle$ denote orthonormal states for subsystem $A(B)$ and
$p_k\geq 0$, $\sum_{k=1}^{n_s} p_k=1$,  allows to easily distinguish separable
pure states ($n_s=1$) from entangled states ($n_s\geq 2$). Here $n_s$ is the
Schmidt rank of $|\Psi_{AB}\rangle$ ($n_s\leq {\rm Min}[d_A,d_B]$, with
$d_{A(B)}$ the Hilbert space dimensions of $A(B)$).  Pure sate entanglement can
be measured by the entanglement entropy \cite{Sch.95}
\begin{equation}E(A,B)=S(\rho_A)=S(\rho_B)=-\sum_{k=1}^{n_s} p_k \log p_k\,, \label{2EE}\end{equation}
where $\rho_{A(B)}={\rm Tr}_{B(A)}\,\rho_{AB}=\sum_{k=1}^{n_s} p_k
|k_{A(B)}\rangle\langle k_{A(B)}|$, with $\rho_{AB}=|\Psi_{AB}\rangle\langle
\Psi_{AB}|$, are the reduced states of $A(B)$ and $S(\rho)=-{\rm
Tr}\,\rho\log\rho$ is the von Neumann entropy.  We  will set in what follows $\log p\equiv
\log_2\,p$, such that $E(A,B)=1$ for a maximally entangled two-qubit state
($n_s=2$, $p_1=p_2=1/2$).

On the other hand, a general mixed state $\rho_{AB}$ ($\rho_{AB}\geq 0$, ${\rm
Tr}\,\rho_{AB}=1$)  of a bipartite system $A+B$ is {separable} iff it can be
expressed as a {\it convex mixture} of product states \cite{RF.89}:
\begin{equation}\rho_{AB}\;{\rm separable}\;\Leftrightarrow\rho_{AB}
=\sum_\alpha p_\alpha
\rho_A^\alpha\otimes\rho_B^\alpha\,,\;\;p_\alpha>0\,,\label{3s}\end{equation}
where $\sum_\alpha p_\alpha=1$ and $\rho_{A(B)}^\alpha$ denote mixed states for
subsystem $A\,(B)$.  Otherwise, it is  entangled. The meaning is that a
separable state can be created by LOCC,  i.e.,  Alice prepares a state
$\rho_A^\alpha$ with probability $p_\alpha$ and tells Bob to prepare a partner
state $\rho_B^\alpha$.

For pure states  $\rho_{AB}=|\Psi_{AB}\rangle\langle \Psi_{AB}|$,  Eq.\
(\ref{3s}) is equivalent to the previous definition
($|\Psi_{AB}\rangle=|\Psi_A\rangle|\Psi_B\rangle$), but  in the case of
mixed states, product states $\rho_{AB}=\rho_A\otimes \rho_B$ are just a very
particular case of separable states. The latter also include:  a) {\it
classically correlated states}, i.e. states diagonal in a standard product
basis $\{|ij\rangle\equiv|i_A\rangle|j_B\rangle\}$,
\begin{equation}\rho_{AB}=
\sum_{i,j}p_{ij}|i_A\rangle\langle i_A|\otimes |j_B\rangle \langle
 j_B|\,,\;\;p_{ij}\geq 0\,,\label{4cc}\end{equation}
where  $\sum_{i,j}p_{ij}=1$ and $|i_{A(B)}\rangle$ are orthonormal states of
$A(B)$, b) {\it classically correlated states from one of the subsystems},  say
$B$,  which are of the form
\begin{equation}\rho_{AB}=\sum_{j}p_{j}\rho_{A/j}\otimes |j_B\rangle\langle j_B|\,,\;\;p_j\geq 0\,,
 \label{5cc}\end{equation}
where $\sum_j p_j=1$ and $\rho_{A/j}$ are states of $A$, and which are then
diagonal in a {\it conditional} product basis
$\{|i_{j}j\rangle\equiv|i_{A/j}\rangle|j_B\rangle\}$ with $|i_{A/j}\rangle$ the
eigenstates of $\rho_{A/j}$ (the case (\ref{4cc}) recovered when all
$\rho_{A/j}$ commute), and c) convex mixtures of product states which are not of the previous
forms a) or b). The latter typically possess entangled eigenstates. For this
reason, it is much more difficult to determine whether a mixed state is
separable or entangled. The well known positive partial transpose criterion
\cite{PE,HO} ($\rho^{t_B}_{AB}\geq 0$, with $\rho^{t_B}_{ij,kl}=\rho_{il,kj}$
for $\rho_{ilkj}=\langle il|\rho_{AB}|kj\rangle$) provides a necessary
criterion for separability, which is sufficient for two-qubit or qubit-qutrit
states.

For  mixed states, the marginal entropies $S(\rho_A)$, $S(\rho_B)$ no longer
provide a measure of entanglement. Instead, it is possible to use the
entanglement of formation \cite{EF}, defined through the convex roof extension
of the pure state definition:
\begin{equation} E(A,B)=\mathop{\rm Min}_{\sum_\alpha p_\alpha |\Psi_{AB}^\alpha\rangle
\langle\Psi_{AB}^\alpha|=\rho_{AB}}S(\rho_{A}^\alpha)\,,\label{6Ef}\end{equation}
where the minimization is over all decompositions of $\rho_{AB}$ as convex
mixtures of pure states ($p_\alpha\geq 0$, $\sum_\alpha p_\alpha=1$) and
$S(\rho_A^\alpha)=S(\rho_B^\alpha)$  is the entanglement entropy of the pure
state $|\Psi_{AB}^\alpha\rangle$. Eq.\ (\ref{6Ef}) vanishes iff $\rho_{AB}$ is
separable, and reduces to the entanglement entropy (\ref{2EE}) for pure states.
It is an entanglement monotone \cite{EM}, i.e.,  it does not increase by LOCC,
staying unaltered under local unitary operations $\rho_{AB}\rightarrow
U_A\otimes U_B\,\rho_{AB}\,U^\dagger_A\otimes U^\dagger_B$. Its evaluation is,
however, difficult in general. A general analytic expression has been derived
just for the two-qubit case \cite{WW.98}, which will be specified in sec.\ 3.

While the marginal entropies are no longer entanglement indicators, it can
still be shown \cite{HO.96} that if $S(\rho_{A})>S(\rho_{AB})$ or
$S(\rho_B)>S(\rho_{AB})$, $\rho_{AB}$ is entangled, i.e.,
\begin{equation}\rho_{AB}\;{\rm separable}\;\Rightarrow\;S(\rho_A)\leq S(\rho_{AB})
\,,\;\;\;S(\rho_B)\leq S(\rho_{AB})\,. \label{S}\end{equation}
Eq.\ (\ref{S}) provides an {\it entropic criterion for separability}
\cite{HO.96} (necessary but not sufficient in general), which can be also
extended to more general entropic forms \cite{NC.02,RC.02}  and which will be
invoked in sec.\ 2.3.

\subsection{Quantum Discord}

For the classically correlated states (\ref{4cc}) or in general (\ref{5cc}),
there is a complete local measurement on $B$  which leaves the state unaltered.
This is not the case for entangled states nor for separable states not of the
form (\ref{4cc}) or (\ref{5cc}). Let us recall here that a general positive
operator valued measurement (POVM) \cite{NC.00}) on system $A+B$ is defined by
a set of operators $\{M_j\}$ satisfying $\sum_j M_j^\dagger M_j=I_{AB}\equiv
I_A\otimes I_B$, such that the probability of outcome $j$ and the joint state
after such outcome are
\begin{equation} p_j={\rm Tr}\,\rho_{AB}\,M_j\,,\;\;\;\rho'_{AB/j}=M_j
\rho_{AB}M_j^\dagger/p_j\,.\label{pm}\end{equation}
The post-measurement state if the outcome is unknown is then
\begin{equation}\rho'_{AB}=\sum_j p_j \rho'_{AB/j}=
 \sum_j M_j\rho_{AB}M_j^\dagger\,.\label{rhop}\end{equation}
Standard projective measurements correspond to the case where the $M_j$ are
orthogonal projectors ($M_k M_j=\delta_{jk}M_j$), while a local measurement on
$B$ corresponds to $M_j=I_A\otimes M_j^B$. By a complete local measurement on
$B$ we will mean  one based on rank one orthogonal projectors $M_j^B=P_j^B$. It
is then apparent that the states (\ref{4cc}) and (\ref{5cc}) remain unchanged
($\rho'_{AB}=\rho_{AB}$) after a local measurement on $B$ based on the
projectors $P_j^{B}=|j_{B}\rangle\langle j_{B}|$. For the states (\ref{4cc})
(but not necessarily (\ref{5cc})) there is also a local measurement on $A$
(that based on the projectors $|i_A\rangle\langle i_A|$) which leaves them
unchanged.

The quantum discord \cite{Zu.01,HV.01,Zu.03} is a measure of quantum
correlations which, unlike the entanglement of formation, can distinguish the
classically correlated states (\ref{5cc}) from the rest of separable states: It
vanishes iff $\rho_{AB}$ is of the form (\ref{4cc}) or (\ref{5cc}), being
positive in the other separable states c), and reduces to the entanglement
entropy (\ref{2EE}) in the case of pure states. It can be defined as the
minimum difference between two distinct quantum versions of the mutual
information, or equivalently, of the conditional entropy:
\begin{eqnarray}D(A|B)&=&\mathop{\rm Min}_{M_B}[I(A,B)-I(A,B_{M_B})]\nonumber\\&=&
\mathop{\rm Min}_{M_B} S(A|B_{M_B})-S(A|B)\,,\label{D2}\end{eqnarray}
where the minimization is over all local measurements $M_B$ on $B$ and
\begin{eqnarray}I(A,B)&=&S(\rho_A)-S(A|B)\,,\nonumber\\
S(A|B)&=&S(\rho_{AB})-S(\rho_B)\,,
\label{SAcB}\end{eqnarray}
are, respectively, the standard quantum mutual information and conditional entropy while
\begin{eqnarray}I(A,B_{M_B})&=&S(\rho_A)-S(A|B_{M_B})\,,\nonumber\\
 S(A|B_{M_B})&=&\sum_ j p_j S(\rho_{A/j})\,,\label{SAcB2}\end{eqnarray}
are the mutual information and conditional entropy after the local measurement
$M_B$, with  $\rho_{A/j}={\rm Tr}_B\,\rho'_{AB/j}$ the reduced state of $A$
after outcome $j$. Eq.\ (\ref{D2}) is always non-negative \cite{Zu.01,HV.01}, a
property which arises from the concavity of the {\it conditional} von Neumann
entropy \cite{Whl}.

In the case of complete local projective measurements $M_B$ we have
\begin{equation}S(A|B_{M_B})=S(\rho'_{AB})-S(\rho'_B)\,,\label{SACB3}\end{equation}
where $\rho'_B={\rm Tr}_A\,\rho'_{AB}$ and $\rho'_{AB}$ is the post-measurement
state (\ref{rhop}). It is then apparent that if the state is of the form
(\ref{4cc}) or (\ref{5cc}), a measurement $M_B$ based on the projectors
$P_j^B=|j_B\rangle \langle j_B|$ leads to $S(A|B_{M_B})=S(A|B)$ and hence
$D(A|B)=0$. For all other states (i.e., entangled states or separable states
not of the form (\ref{4cc}) or (\ref{5cc})), $D(A|B)>0$. In the case of pure
states, $S(\rho_{AB})=0$ while $S(A|B_{M_B})=0$ if $M_B$ is any complete local
measurement, entailing $D(A|B)=S(\rho_B)=E(A,B)$. For mixed states,
the quantum discord can be related to the entanglement of formation $E(A,C)$
with a third system $C$ purifying the whole system \cite{KW,MD.11,CAB.11,Fan}.

The mutual information $I(A,B)$ is a measure of all correlations between $A$
and $B$, being non-negative and vanishing just for product states
$\rho_{AB}=\rho_A\otimes\rho_B$. The bracket in  (\ref{D2}) can then be
interpreted as the difference between all correlations (classical+quantum)
present in the original state minus the classical correlations left after the
local measurement on $B$, which leaves then the quantum correlations. The
evaluation of Eq.\ (\ref{D2}) is, nevertheless, difficult in the general case,
being in fact an NP complete problem \cite{NP} due to the minimization over
all possible local measurements $M_B$. Nonetheless, the minimum is always
attained for measurements based on rank one projectors $P_j^B$, not necessarily
orthogonal \cite{Mo.11,GR.14}.

\subsection{Information Deficit}

The one-way information deficit can be considered as an alternative  measure of
quantum correlations, with basic properties similar to those of the quantum
discord. It can be defined as \cite{HHH.05,SKB.11,Zu.03,RCC.10}
\begin{equation}
I(A|B)=\mathop{\rm Min}_{M_B} S(\rho'_{AB})-S(\rho_{AB})\,,\label{ID}
\end{equation}
where $\rho'_{AB}$ is the post-measurement state (\ref{rhop}) and $M_B$ is here
restricted to complete local projective measurements on $B$, such that
$\rho'_{AB}$ is of the form (\ref{5cc}). Like the quantum discord, Eq.\
(\ref{ID}) is a non-negative quantity which also vanishes just for the states
(\ref{4cc}) or (\ref{5cc}),  and which also reduces to the entanglement entropy
(\ref{2EE}) in the case of pure states. These properties will be shown below
in a more general context, although they are also apparent
from the alternative expression
\begin{equation}
I(A|B)=\mathop{\rm Min}_{M_B} S(\rho_{AB}||\rho'_{AB})\,,\label{IDr}
\end{equation}
where $S(\rho||\sigma)={\rm Tr}\,\rho(\log\rho-\log\sigma)$ is the relative
entropy \cite{Whl,Ve.02}, a quantity satisfying $S(\rho||\sigma)\geq 0$, with
$S(\rho||\sigma)=0$ iff $\rho=\sigma$. Eq.\ (\ref{IDr}) can be shown by noting
that $\rho'_{AB}$ is the diagonal part of $\rho_{AB}$ in the basis defined by
the projective measurement (the minimization in (\ref{IDr}) can in fact be
extended to all $\rho'_{AB}$ of the form (\ref{5cc}) \cite{RCC.10}).
Nevertheless, differences with the quantum discord may arise in the minimizing
measurement, as discussed in the next section. We also note  that if the
minimizing measurement of $D(A|B)$ is projective and in the basis of
eigenstates of $\rho_B$, then $\rho'_B=\rho_B$ and Eqs.\
(\ref{D2})--(\ref{SACB3}) lead to $D(A|B)=I(A|B)$. Otherwise $D(A|B)\leq
I(A|B)$, since for projective measurements Eqs.\ (\ref{D2})--(\ref{SACB3}) imply
$D(A|B)\leq S(\rho'_{AB})-S(\rho_{AB})-[S(\rho'_B)-S(\rho_B)]\leq
S(\rho'_{AB})-S(\rho_{AB})$.

Eq.\ (\ref{ID}) admits a simple interpretation in terms of the entanglement
generated between the system and a measuring apparatus $M$ performing the
complete local measurement \cite{SKB.11}. The measurement on the local basis
$\{|i_B\rangle\}$ can be represented through a unitary operator $U_{BM}$
satisfying $U_{BM}|j_B 0_M\rangle=|j_B j_M\rangle$, where $|0_M\rangle$ is the
initial state of the apparatus and $\{|j_M\rangle\}$ an orthogonal basis of
$M$,  such that
\begin{eqnarray}\rho'_{AB}&=&{\rm Tr}_M\,\rho'_{ABM}\,,\nonumber\\
\rho'_{ABM}&=&(I_A\otimes U_{BM})
(\rho_{AB}\otimes |0_M\rangle\langle 0_M|)(I_A\otimes
U_{BM}^\dagger)\,.\;\;\;\label{oper}\end{eqnarray} Since
$S(\rho_{AB})=S(\rho_{AB}\otimes |0_M\rangle\langle 0_M|)=S(\rho'_{ABM})$, it
is seen that Eq.\ (\ref{ID}) is the difference between the entropy of the
subsystem $AB$ and that of the total system $ABM$ after the measurement, and``
according to Eq.\ (\ref{S}), such difference can be positive only if there is
entanglement between $AB$ and $M$. Thus, a positive $I(A|B)$ indicates that
entanglement between $AB$ and $M$ is generated by {\it any} complete local
measurement $M_B$. On the other hand, if $I(A|B)=0$, then $\rho_{AB}$ is of the
form (\ref{5cc}) and for a measurement in the basis $\{|j_B\rangle\}$,
$\rho'_{ABM}=\sum_{j}p_j\rho_{A/j}\otimes |j_Bj_M\rangle\langle j_B j_M|$ is
clearly separable,  so that no entanglement is generated by this measurement.
It can be shown \cite{SKB.11} that Eq.\ (\ref{ID}) coincides in fact with the
minimum distillable entanglement between $AB$ and $M$ generated by the complete
local measurement on $B$. A similar interpretation for the quantum discord in
terms of  the minimum partial distillable entanglement can also be obtained
\cite{SKB.11}. Other operational interpretations can be found in
\cite{AD.12,MD.11,CAB.11,Fan,Na,Aex}.

\subsection{Generalized Information Deficit}

It is possible in principle to extend Eq.\ (\ref{ID}) to more general entropic
forms, since in contrast with the quantum discord (\ref{D2}), its positivity is
not related to specific properties of the von Neumann entropy $S(\rho)$, as shown below.  We
consider here generalized entropies of the form \cite{CR.02}
\begin{equation}S_f(\rho)={\rm Tr}\,f(\rho)\,,\label{Sf}\end{equation}
where ${\rm Tr}\,f(\rho)=\sum_i f(p_i)$, with $p_i$ the eigenvalues of $\rho$ and
$f(p)$ a smooth strictly concave real function defined for $p\in[0,1]$ and satisfying
$f(0)=f(1)=0$. These entropies fulfill the same basic properties as the von
Neumann entropy,  with the exception of additivity: We have   $S_f(\rho)\geq
0$, with $S_f(\rho)=0$ iff $\rho$ is a pure state ($\rho^2=\rho$), while all
$S_f(\rho)$ are maximum for the maximally mixed state $\rho=I/d$, where $d={\rm
Tr}\,I$ is the Hilbert space dimension of the system.  Moreover, they are
strictly concave, i.e., $S_f(\sum_\alpha p_\alpha \rho_\alpha)\geq \sum_\alpha
p_\alpha S_f(\rho_\alpha)$, for $p_\alpha>0$, $\sum_\alpha p_\alpha=1$, with
equality iff all $\rho_\alpha$ are coincident.  The von Neumann entropy is
obviously recovered for $f(\rho)=-\rho\log\rho$.

Concavity of $S_f(\rho)$ implies  the fundamental majorization property
\begin{equation}
\rho'\prec\rho\Rightarrow S_f(\rho')\geq S_f(\rho)\,,\label{prec}\end{equation}
where $\rho'\prec\rho$ indicates that $\rho'$ is {\it majorized} by  $\rho$
\cite{Bha,Whl} (also denoted as $\rho'$ {\it more mixed} than $\rho$):
\begin{equation}\rho'\prec\rho\Leftrightarrow\;\sum_{j=1}^ip'_j
\leq \sum_{j=1}^i p_j\,,\;\;i=1,\ldots,d-1\,,
 \label{prec0}\end{equation}
where $p_j$, $p'_j$ denote the eigenvalues of $\rho$ and $\rho'$ sorted in {\it
decreasing} order (equality in (\ref{prec0}) obviously holds for $i=d$). If the
dimensions of $\rho$ and $\rho'$ differ, Eq.\ (\ref{prec}) still holds (for
$f(0)=0$) after completing with zeros the smallest set of eigenvalues.
Conversely, while the reverse of Eq.\ (\ref{prec}) does not necessarily hold,
indicating that majorization provides a more strict concept of mixedness or
disorder than that defined by a single choice of entropy,  it does hold if
$S_f(\rho')\geq S_f(\rho)$ $\forall$ $f$ of  the previous form \cite{RC.02}:
\begin{equation}S_f(\rho')\geq S_f(\rho)\;\forall\;S_f\;
\Rightarrow \rho'\prec\rho\label{prec2}\,.\end{equation}
Eq.\ (\ref{prec}) remains actually valid for more general entropic forms (like
increasing functions $F(S_f)$ of $S_f$ or in general, Schur concave functions
\cite{Bha}), but Eq.\ (\ref{prec2}) indicates that the forms (\ref{Sf}) are
already sufficient to capture majorization. Among the various properties
implied by majorization, we mention that for states with the same dimension,
$\rho'\prec\rho$ iff $\rho$ is a convex mixture of unitary transformations of
$\rho$ \cite{Bha,Whl}, i.e., iff  $\rho'=\sum_\alpha p_\alpha U_\alpha\rho
U_\alpha^\dagger$,  with $U_\alpha$ unitary and $p_\alpha\geq 0$.

Now, for any projective measurement (local or non-local) performed on the
system $A+B$, it can be easily shown that $S_f(\rho'_{AB})\geq S_f(\rho_{AB})$
$\forall$ $S_f$, i.e.,
\begin{equation} \rho'_{AB}\prec\rho_{AB}\,.\label{prec4}\end{equation}
The reason is that the post measurement state $\rho'_{AB}$ conserves just the
diagonal elements $p'_\nu=\langle \nu'|\rho_{AB}|\nu'\rangle$ of  $\rho_{AB}$
in a certain orthonormal basis $\{|\nu'\rangle\}$ determined  by the projectors
and hence, $S_f(\rho'_{AB})=\sum_\nu f(p'_\nu)=\sum_\nu f(\sum_\mu|\langle
\mu|\nu'\rangle|^2 p_\mu)\geq \sum_{\mu,\nu}|\langle
\mu|\nu'\rangle|^2f(p_\mu)=S_f(\rho_{AB})$, where $p_\mu$ and $|\mu\rangle$
denote here the eigenvalues and eigenvectors of $\rho_{AB}$.  This relation
is not restricted to rank one projectors (just choose an orthonormal basis
$\{|\nu'\rangle\}$ where $\rho'_{AB}$ is diagonal), so that  it holds for local
projective measurements. Eq.\ (\ref{prec4}) remains actually valid for any
measurement satisfying $\sum_j M_j M_j^\dagger=I_{AB}$, i.e., which leaves the
maximally mixed state $I_{AB}/d_{AB}$ unchanged \cite{RCC.10}.

Note also that strict concavity of $S_f$ implies
$S_f(\rho'_{AB})=S_f(\rho_{AB})$ iff $\rho'_{AB}=\rho_{AB}$, as is apparent
from the previous demonstration. In fact, if the off diagonal elements of
$\rho_{AB}$ in the measured basis are sufficiently small,  a second order
expansion of $S_f(\rho_{AB})$ leads to \cite{RCC.10}
\begin{equation}S_f(\rho'_{AB})-S_f(\rho_{AB})\approx \sum_{\mu<\nu}
\frac{f'(p'_\mu)-f'(p'_\nu)}{p'_\nu-p'_\mu}|\langle \nu'|\rho_{AB}|\mu'\rangle|^2\,,
\label{norm}\end{equation}
where the fraction is always positive due to the strict concavity of $f$ (and
should be replaced by its limit $-f''(p'_\mu)$ if $p'_\nu\rightarrow p'_\mu$).
Eq.\ (\ref{norm}) is essentially the square of a weighted norm of the
off-diagonal elements of $\rho_{AB}$ in the measured basis (i.e., of those lost
in the measurement), and is therefore non-negative, vanishing (if $f''(p)<0$
$\forall$ $p\in(0,1)$) only if all off-diagonal elements are zero.

We may then define the quantity \cite{RCC.10,RCC.11}
\begin{equation}I_f(A|B)=\mathop{\rm Min}_{M_B} S_f(\rho'_{AB})-S_f(\rho_{AB})\label{If}\,,\end{equation}
where the minimization is again over all complete local measurements on $B$.
Eq.\ (\ref{If}) is non-negative, due to Eq.\ (\ref{prec4}), and vanishes iff
$\rho'_{AB}=\rho_{AB}$, i.e., iff $\rho_{AB}$ is already of the classically
correlated form (\ref{4cc}) or (\ref{5cc}). It therefore vanishes only for the
states with zero quantum discord. It obviously also remains invariant under
local unitary operations.

In the case of pure states, it can be shown \cite{RCC.10} that the minimum of
Eq.\ (\ref{If}) is always attained for a measurement in the basis
$\{|k_B\rangle\}$ determined by the Schmidt decomposition (\ref{1SD}), i.e., in
the basis formed by the eigenstates of $\rho_B$, which leads to
\begin{equation} I_f(A|B)=S_f(\rho_A)=S_f(\rho_B)=\sum_{k=1}^{n_s}
f(p_k)\,,\;\;\;(\rho_{AB}\;{\rm pure})\,.\label{Ifpure}\end{equation}
It therefore reduces to the {\it generalized entanglement entropy}
$S_f(\rho_A)=S_f(\rho_B)$ of the pure state. The entanglement entropy can then
be identified with the minimum information loss due to a local measurement
\cite{RCC.10}. It is apparent that for pure states, $I_f(A|B)=I_f(B|A)$, a
property which does not hold in the general case.

In the case of the von Neumann entropy, $I_f(A|B)$ becomes the standard
information deficit (\ref{ID}) and Eq.\ (\ref{Ifpure}) implies that for pure
states, it will coincide with the standard (von Neumann) entanglement entropy,
like the quantum discord. Nevertheless, an important difference arises in the
minimizing measurement,  since  that for the latter becomes undetermined in the
case of pure states (it can be any measurement based on rank one projectors
\cite{GR.14}), whereas  all $I_f(A|B)$, including $I(A|B)$, require a
measurement in the basis $\{|k_B\rangle\}$, which is fully undetermined only in the
case of maximally mixed marginals.

Like the standard information deficit, $I_f(A|B)$ is also an indicator of the
minimum entanglement between the system and the measurement apparatus $M$
generated by a complete local measurement. The von Neumann entropic criterion
for separability (\ref{S}) can actually be extended to any $S_f$ \cite{RC.02}:
 \begin{equation}\rho_{AB}\;{\rm separable}\;\Rightarrow\;S_f(\rho_A)
\leq S_f(\rho_{AB})\,,\;\;\;S_f(\rho_B)\leq
 S_f(\rho_{AB})\;.\label{Sfg}\end{equation}
The validity of Eq.\ (\ref{Sfg}) for all $S_f$ is stronger than the von Neumann
based criterion (\ref{If}) \cite{RC.02}, and equivalent to the disorder
criterion of separability \cite{NC.02} ($\rho_{AB}\;{\rm separable}$
$\Rightarrow$ $\rho_{AB}\prec\rho_{A(B)}$). By the same arguments given below
Eq.\ (\ref{oper}), it follows that a positive $I_f(A|B)$, i.e.,
$S_f(\rho'_{AB})>S_f(\rho_{AB})=S_f(\rho'_{ABM})$, is indicating the existence
of entanglement between $AB$ and $M$ after {\it any} complete local projective
measurement on $B$.

\subsection{Minimizing measurement}

Eq.\ (\ref{Ifpure}) reflects an universal property exhibited by the local
measurement minimizing $I_f(A|B)$ for pure states:  It is the same for all
$S_f$. Such measurement, i.e., a  measurement in the basis $\{|k_B\rangle\}$
determined by the Schmidt decomposition of the pure state,  is also optimum,
for {\it all} $S_f$, for the mixture of the pure state with the maximally mixed
state \cite{RCC.10},
\begin{equation} \rho_{AB}=q|\Psi_{AB}\rangle\langle \Psi_{AB}|+(1-q)I_{AB}/d_{AB}\,,
\;\;q\in[0,1]\,.\label{mix}\end{equation}

These states exhibit then an unambiguous {\it least disturbing local
measurement}, in the sense that it minimizes all $I_f(A|B)$ and leads to a
``least mixed'' post-measurement state
\[\rho'_{AB}=q\sum_{k=1}^{n_s}p_k|k_A\rangle\langle k_A|\otimes |k_B\rangle\langle k_B|
 +(1-q)I_{AB}/d_{AB}\,,\]
which {\it majorizes} any other post-measurement state emerging after a local
measurement. This property does not hold for an arbitrary initial state
$\rho_{AB}$.

In the general case, the projective measurement $M_B=\{|j_B\rangle\langle
j_B|\}$ minimizing $I_f(A|B)$ may depend on the choice of entropy $S_f$. It can
be shown that it  must satisfy the necessary stationary condition \cite{RCC.11}
\begin{equation}{\rm Tr}_A[f'(\rho'_{AB}),\rho_{AB}]=0\,,\label{stat}\end{equation}
where $f'$ denotes the derivative of $f$ and $\rho'_{AB}$ is the
post-measurement state (\ref{rhop}). Eq.\ (\ref{stat}) implies, explicitly,
$\sum_i [f'(p'_{ij})\langle i_j j|\rho_{AB}|i_jk\rangle-f'(p'_{ik})\langle i_k
j|\rho_{AB}|i_k k\rangle]=0$, where $p'_{ij}=\langle
i_jj|\rho_{AB}|i_jj\rangle$ and $|i_jj\rangle=|i_{A/j}\rangle|j_B\rangle$, with
$|i_{j/A}\rangle$ the eigenstates of $\rho_{A/j}$. The minimizing measurement
basis will not coincide in general with the eigenstates of $\rho_B$, even
though this holds for certain states, like  pure states and the mixtures
(\ref{mix}). Eq.\ (\ref{stat}) shows that the eigenstates of $\rho_B$ will be
stationary for any state $\rho_{AB}$ where the non-zero off-diagonal
elements are of the form $\langle ij|\rho_{AB}|kl\rangle$ with $i\neq k$ {\it and} $j\neq
l$, where $|ij\rangle\equiv |i_A\rangle |j_B\rangle$ and $|i_A\rangle$,
$|j_B\rangle$ are the eigenstates of $\rho_A$ and $\rho_B$ respectively
\cite{RCC.11}.

In the case of the quantum discord, and for $M_B$ restricted to complete local
projective measurements, Eq.\ (\ref{stat}) is to be replaced by (here
$f'(\rho)=-\log\rho$) \cite{RCC.11}
\begin{equation}{\rm Tr}_A[f'(\rho'_{AB}),\rho_{AB}]-[f'(\rho'_{AB}),\rho_B]=0\,.
\label{statqd}\end{equation}
More explicit expressions can be obtained for a two-qubit system, where we may
write a general state as
\begin{equation}\rho_{AB}={\textstyle\frac{1}{4}}(I_{AB}+\bm{r}_A\cdot\bm{\sigma}_A+
\bm{r}_B\cdot\bm{\sigma}_B+\bm{\sigma}_A^tJ\bm{\sigma}_B)\,,
 \label{stat2q}\end{equation}
where $\bm{\sigma}_{A}=\bm{\sigma}\otimes I$, $\bm{\sigma}_{B}=I\otimes\bm{\sigma}$,
with $\bm{\sigma}^t=(\sigma_x,\sigma_y,\sigma_z)$ the Pauli operators, and
$I_{AB}=I\otimes I$ the identity. Since ${\rm Tr}\,\sigma_\mu=0$ and ${\rm
Tr}\,\sigma_{\mu}\sigma_\nu=2\delta_{\mu\nu}$ for $\mu,\nu=x,y,z$, we have
($\langle O\rangle\equiv{\rm Tr}\,\rho_{AB}\,O)$
\begin{equation}\bm{r}_{A(B)}=\langle \bm{\sigma}_{A(B)}\rangle,\;\;J=
\langle \bm{\sigma}_A\bm{\sigma}_B^t\rangle\,.\label{vm}\end{equation}
A complete projective measurement on $B$ corresponds to a spin measurement
along the direction of a unit vector $\bm{k}$, represented by projectors
$P_{\pm\bm{k}}=\frac{1}{2}(I\pm\bm{k}\cdot\bm{\sigma})$. After this
measurement, Eq.\  (\ref{stat2q}) becomes
\begin{equation}\rho'_{AB}={\textstyle\frac{1}{4}}[I+\bm{r}_A\cdot\bm{\sigma}_A+
(\bm{r}_B\cdot \bm{k})\bm{k}\cdot\bm{\sigma}_B+(\bm{\sigma}_A^tJ\bm{k})\bm{k}\cdot\bm{\sigma}_B]\,.
\label{rhopk}\end{equation}
Eq.\ (\ref{stat}) leads then to the explicit equation \cite{RCC.11}
\begin{equation}\alpha_1 \bm{r}_B+\alpha_2 J^t\bm{r}_A+\alpha_3 J^tJ\bm{k}=\lambda \bm{k}\,,
\label{stk}\end{equation} where
$(\alpha_1,\alpha_2,\alpha_3)=\frac{1}{4}\sum_{\mu,\nu=\pm
1}f'(p'_{\mu\nu})(\nu,\frac{\mu\nu}{|\bm{r}_A+J\bm{k}|},\frac{\mu}{|\bm{r}_A+J\bm{k}|})$,
$p'_{\mu\nu}=\frac{1}{4}(1+\nu\bm{r}_B\cdot\bm{k}+\mu|\bm{r}_A+\nu J\bm{k}|)$
are the eigenvalues of $\rho'_{AB}$, with $\mu,\nu=\pm 1$, and $\lambda$ is a
proportionality factor. In the case of the quantum discord, Eq.\ (\ref{statqd})
leads to a similar equation, with $f(p)\rightarrow -p\log p$ and
$\alpha_1\rightarrow \alpha_1-\frac{1}{2}\log p'_-/p'_+$, where
$p'_{\pm}=\frac{1}{2}(1+\bm{r}_B\cdot\bm{k})$ are the eigenvalues of $\rho'_B$
\cite{RCC.11}.

\subsection{Particular cases}

One of the advantages of the generalized information deficit (\ref{If}) is the
possibility of using simple entropic forms which can be more easily evaluated
(and measured) than the von Neumann entropy. For instance, if
$f(\rho)=2(\rho-\rho^2)$, Eq.\ (\ref{Sf}) becomes the so called linear entropy
\begin{equation}S_2(\rho)=2(1-{\rm Tr}\,\rho^2)\label{S2}\,,\end{equation}
which follows from the linear approximation $\ln\rho\approx \rho-I$ in the von
Neumann entropy, but is actually a quadratic function of $\rho$, i.e., a linear
function of the purity $P(\rho)={\rm Tr}\,\rho^2$. It is the simplest entropic
form and its evaluation does not require the knowledge of the eigenvalues of
$\rho$ (see Eq.\ (\ref{S2tq}) below). Moreover, purity, and hence $S_2(\rho)$,
can be experimentally determined without a full state tomography \cite{Exp}.
Eq.\ (\ref{S2}) is actually the $q=2$ case of the Tsallis entropies
\cite{Ts.09}, obtained for $f(\rho)=\frac{\rho-\rho^q}{1-2^{1-q}}$:
\begin{equation} S_q(\rho)=\frac{1-{\rm Tr}\,\rho^q}{1-2^{1-q}}\; ,\,\,\,\,\,\,
 q>0\;.\label{TS}\end{equation}
Eq.\ (\ref{TS}) approaches the von Neumann entropy $S(\rho)$ for $q\rightarrow
1$, being strictly concave for $q>0$. We have normalized (\ref{S2}) and
(\ref{TS}) such that $S_q(\rho)=1$ for a maximally mixed two-qubit state.

In the case (\ref{S2}), it is first seen that for post-measurements states $\rho'_{AB}$,
\begin{equation} S_2(\rho'_{AB})-S_2(\rho_{AB})=2\,{\rm Tr}
\,(\rho_{AB}^2-{\rho'}^2_{AB})=2||\rho_{AB}'-\rho_{AB}||^2\,,\end{equation}
where $||O||^2={\rm Tr}\,O^\dagger O$. Hence, the local projective measurement minimizing
$S_2(A|B)$, which is that  maximizing the post-measurement purity
$P(\rho'_{AB})$, leads to the post-measurement state with the minimum Hilbert-Schmidt
distance  to the original state. The associated deficit
\begin{equation}I_2(A|B)=\mathop{\rm Min}_{M_B}S_2(\rho'_{AB})-S_2(\rho_{AB})\,,\label{I2}\end{equation}
coincides, apart from a constant factor, with the geometric discord
\cite{DVB.10, Mo.11, RCC.10}. For pure states,  $I_2(A|B)$ will then coincide
with the linear marginal entropies:
\begin{equation} I_2(A|B)=S_2(\rho_A)=S_2(\rho_B)=2(1-\sum_{k=1}^{n_s}p_k^2)\,.\label{I2p}\end{equation}
In two qubit systems, Eq.\ (\ref{I2p}) is just the squared  {\it concurrence}
\cite{WW.98} of the pure state $\rho_{AB}$.

While as a measure the geometric discord fails to satisfy some additional
properties fulfilled by the quantum discord or the information deficit \cite{Pia}, it
offers the enormous advantage of a simple analytic evaluation in qudit-qubit
systems \cite{DVB.10,RCC.11,VR.11}, as discussed below, also admitting through the
purity a more direct experimental access. Moreover, Eq.\ (\ref{norm}) shows
that if $\rho_{AB}$ is close to the maximally mixed state $I_{AB}/d_{AB}$, all $I_f(A|B)$ will
become proportional to $I_2(A|B)$ \cite{RCC.10}, as in this case
$\frac{f'(p'_\mu)-f'(p'_\nu)}{p'_{\nu}-p'_{\mu}}\approx -f''(\frac{1}{d_{AB}})$ is nearly constant.
In fact, all $S_f(\rho)$ are linearly related to $S_2(\rho)$ in this limit \cite{GR.14}.

Any  state of a general system $A+B$ can be written in the form
(\ref{stat2q}), replacing the Pauli operators by a complete set of orthogonal
operators $\bm{\sigma}$ in $A$ and $B$  satisfying ${\rm Tr}\,\sigma_{\mu}=0$,
${\rm Tr}\,\sigma_{\mu}\sigma_{\nu}=d_{A(B)}\delta_{\mu\nu}$:
\begin{equation}\rho_{AB}={\textstyle\frac{1}{d_A d_B}}(I_{AB}+
\bm{r}_A\cdot\bm{\sigma}_A+\bm{r}_B\cdot\bm{\sigma}_B+\bm{\sigma}_A^tJ\bm{\sigma}_B)\,,
 \label{statgq}\end{equation}
where $\bm{r}_{A(B)}$ and $J$ (now a $d_A\times d_B$ matrix) are again given by
Eq.\ (\ref{vm}). The $S_2$ entropy can then be readily evaluated as
\begin{equation}S_2(\rho_{AB})=2[1-{\textstyle\frac{1}{d_A d_B}}(1+|\bm{r}_A|^2+|\bm{r}_B|^2+||J||^2)]\,,
 \label{S2tq}\end{equation}
where $||J||^2={\rm Tr}J^tJ$. If $B$ is now a qubit, the state after a spin
measurement along direction $\bm{k}$  on $B$, will have the form (\ref{rhopk})
with $\frac{1}{4}\rightarrow \frac{1}{2d_A}$. We then obtain, using Eq.\
(\ref{S2tq}),
\begin{equation}S_2(\rho'_{AB})=2-\frac{1}{d_A}(|\bm{r}_A|^2+\bm{k}^t M_2\bm{k})\,,\end{equation}
where $M_2=\bm{r}_B\bm{r}_B^t+J^tJ\,$ is a $\,3\times 3$  positive semidefinite
symmetric matrix. Hence,
$I_2(\bm{k})=S_2(\rho'_{AB})-S_2(\rho_{AB})={\textstyle\frac{1}{d_A}}({\rm
Tr}\, M_2-\bm{k}^tM_2 \bm{k})$.
 Its minimum $I_2(A|B)$ can then be evaluated analytically as \cite{DVB.10,RCC.11}
\begin{equation} I_2(A|B)=\mathop{\rm Min}_{\bm{k}}I_2(\bm{k})={\textstyle\frac{1}{d_A}}
({\rm Tr} M_2-\lambda_1)\label{I2m}\,,
\end{equation}
where $\lambda_1$ is the largest eigenvalue of $M_2$, the minimizing spin
measurement being along the direction of the corresponding eigenvector. Eq.\
(\ref{I2m}) is valid for an arbitrary qudit-qubit state $\rho_{AB}$. Let us
notice that the stationary condition (\ref{stat}) or (\ref{stk}) reduces, for
the linear entropy, precisely to the eigenvalue equation $M_2\bm{k}=\lambda
\bm{k}$, as in this case $f'(\rho'_{AB})\propto \rho'_{AB}$ and hence,
$\alpha_1=\bm{r}_B\cdot\bm{k}$, $\alpha_2=0$ and $\alpha_3=1$ \cite{RCC.11}.
This indicates that the stationary measurements are those along the direction
of the eigenvectors of $M_2$.

For arbitrary $q>0$, we may similarly define the quantities (in what follows
$c_q=1-2^{1-q}$)
\begin{eqnarray}I_q(A|B)&=&\mathop{\rm Min}_{M_B}S_q(\rho'_{AB})-S_q(\rho_{AB})\nonumber\\
&=&\mathop{\rm Min}_{M_B}\,c_q^{-1}{\rm Tr}\,({\rho}_{AB}^q-{\rho'}_{AB}^q)\label{Iq}\,,\\
I^R_q(A|B)&=&\mathop{\rm Min}_{M_B}S^R_q(\rho'_{AB})-S_q^R(\rho_{AB})\nonumber\\
&=&\mathop{\rm Min}_{M_B}\frac{1}{1-q}\log \frac{{\rm Tr}\,{\rho'}_{AB}^q}{{\rm Tr}\,{\rho}_{AB}^q}\label{IRq}\\
&=&\frac{1}{1-q}\log\left[1-\frac{c_q I_q(A|B)}{1-c_q S_q(\rho_{AB})}\right]\,,\label{IRq2}\end{eqnarray}
where
\begin{equation}
S_q^R(\rho)=\frac{1}{1-q}\log\,{\rm Tr}\,\rho^q=\frac{1}{1-q}\log[1-c_q
S_q(\rho)]\,,\;\;q>0\,,\end{equation} are the {\it Renyi} entropies \cite{Whl}, which are
just increasing functions of the Tsallis entropies (\ref{TS}) (and also
approach the von Neumann entropy for $q\rightarrow 1$). Eqs.\
(\ref{Iq})--(\ref{IRq}) are again non-negative, vanishing iff $\rho_{AB}$ is of
the form (\ref{4cc}) or (\ref{5cc}), and approach the von Neumann information
deficit (\ref{ID}) for $q\rightarrow 1$. Eq.\ (\ref{IRq2}) is again just an
increasing function of $I_q(A|B)$ (for fixed $\rho_{AB}$) and does not depend
on the addition of an uncorrelated ancilla $C$ to $A$ ($\rho_{AB}\rightarrow
\rho_{C}\otimes \rho_{AB}$), as ${\rm Tr}\rho_C^q$ cancels out. An analytic
expression for $I_3(A|B)$ valid for any two-qubit state can  also be obtained
\cite{RCC.11}.

\section{Application: Quantum correlations of spin pairs in XY chains}

\subsection{Model and general expressions}

We consider  a spin $1/2$ system with XYZ couplings of arbitrary range,
immersed in a transverse magnetic field $B$ along the $z$ axis.  The
Hamiltonian reads
 \begin{equation}
H=B\sum_{i}\,s_{iz}-{\textstyle\frac{1}{2}}\sum_{\mu=x,y,z} \sum_{i\neq j} J^{ij}_\mu
s_{i\mu}s_{j\mu}\,,
 \label{H}\end{equation}
where $s_{i\mu}$ are the (dimensionless) components of the local spin at site
$i$, and $J^{ij}_{\mu}$ the coupling strengths.

The Hamiltonian (\ref{H}) commutes with the $S_z$ spin parity  operator $P_z$,
irrespective of the coupling range, anisotropy, dimension,  or geometry of the
system \cite{RCM.08,RCM.09},
\begin{equation}
[H,P_z]=0,\;\;P_z=\exp[i\pi\sum_i (s_{iz}+1/2)]=\prod_i(-\sigma_{iz})\,, \label{Pz}
\end{equation}
where $\sigma_{iz}=2s_{iz}$. The non-degenerate eigenstates of $H$ will then
have a definite $S_z$ parity  $P_z=\pm 1$.

Consequently, the reduced density matrix of an arbitrary spin pair  $i,j$ in
any non-degenerate eigenstate $|\Psi_\nu\rangle$, $\rho_{ij}={\rm
Tr}_{\,\overline{(i,j)}}\,|\Psi_\nu\rangle\langle\Psi_\nu|,$ will then commute
with the $S_z$ parity operator of the  pair $P_z^{ij}=\sigma_{iz}\sigma_{jz}$:
$[\rho_{ij},P_z^{ij}]=0$.
In the  standard basis  $\{|00\rangle,|01\rangle,|10\rangle,|11\rangle\}$,
$\rho_{ij}$ will therefore be an $X$-type  state of form
\begin{equation}
\rho_{ij}=\left(\begin{array}{cccc}a_+&0&0&\beta\\0&c_+&\alpha&0\\
 0&\bar{\alpha}&c_-&0\\\bar{\beta}&0&0&a_-\end{array}\right)\,
 \label{rij}\,,\end{equation}
where the coefficients are all real (since $H$ is real in the full standard
basis) and given by ($s_{i\pm}=s_{ix}\pm is_{iy}$)
 \begin{eqnarray}
a_{\pm}&=&{\textstyle\frac{1}{4}}
\pm{\textstyle\frac{1}{2}}\langle s_{iz}+s_{jz}\rangle+\langle s_{iz}s_{jz}\rangle\,,
\label{vm1}\\
c_{\pm}&=&{\textstyle\frac{1}{4}}\pm{\textstyle\frac{1}{2}}\langle s_{iz}-s_{jz}\rangle
-\langle s_{iz}s_{jz}\rangle\,,\label{vm2}\;\;
 (^{\beta}_{\alpha} )=\langle s_{i-}s_{j\mp}\rangle\,,
  \end{eqnarray}
with $a_+ + c_++c_- + a_- = 1$. It corresponds to $\bm{r}_A$ and $\bm{r}_B$
along $z$ in (\ref{stat2q}) ($r_{A(B)}=a_++c_{+(-)}-c_{-(+)}-a_-$),  with  $J$
{\it diagonal}, i.e., $J_{\mu\nu}=4\langle s_{i\mu} s_{j\nu}\rangle=\delta_{\mu\nu} j_\mu$, with
$j_{^x_y}=2(\alpha\pm\beta)$, $j_z=a_++a_--c_+-c_-$. Positivity of $\rho_{ij}$
implies $|\alpha|\leq\sqrt{c_+c_-}$, $|\beta|\leq \sqrt{a_+a_-}$, with
$a_{\pm}$, $c_{\pm}$ non-negative. The single spin density matrix is
\begin{equation}
\rho_i={\rm Tr}_j\,\rho_{ij}=\left(\begin{array}{cc}a_++c_+&0\\0&a_-+c_-\end{array}\right)\,.
\end{equation}
Both $\rho_{ij}$ and $\rho_i$ will obviously be typically mixed  due to the
entanglement with the rest of the chain.

In what follows we will consider translational invariant systems
such that $\langle s_{iz}\rangle$ is site independent, i.e.,
$\langle s_{iz}\rangle=\langle s_{jz}\rangle$  $\forall$ $i,j$, implying
$c_{\pm}=c=\frac{1-a_+-a_-}{2}$. In this situation, $\rho_{ij}=\rho_{ji}$ and
 $D(A|B)=D(B|A)=D$,  $I_f(A|B)=I_f(B|A)=I_f$ $\forall$ $S_f$.

The entanglement of the pair can be measured by the entanglement of
formation (\ref{6Ef}), which for  two qubit states can be evaluated as
\cite{WW.98}
 \begin{equation}
 E=-\sum_{\nu=\pm}q_\nu\log q_\nu
 \,,\;q_{\pm}={\textstyle\frac{1}{2}}(1\pm\sqrt{1-C^2})\,,\label{Eij}
 \end{equation}
where  $C$ is  the concurrence \cite{WW.98}. For the states (\ref{rij}) with $c_{\pm}=c$,
the concurrence  of the pair is given by
\begin{equation}
C_{ij}=2\,{\rm Max}[|\beta|-c,|\alpha|-\sqrt{a_+a_-},0]\,.
 \label{Cij}\end{equation}
The pair entanglement is of parallel type  (as in the Bell states
$\frac{|00\rangle\pm|11\rangle}{\sqrt{2}}$)  if the first entry in  (\ref{Cij})
is positive and  antiparallel  (as in
$\frac{|01\rangle\pm|10\rangle}{\sqrt{2}}$) if the second entry is positive
\cite{Am.06} (just one of them can be positive).

On the other hand, the quantum discord of the pair can be readily evaluated
with the expressions (\ref{rij}) and (\ref{rhopk}) (see \cite{CRC.10} for
details).  The ensuing minimization over the spin measurement direction
$\bm{k}$ (we will consider here just projective measurements) will normally
lead to the direction corresponding to maximum correlation, according to
general arguments of \cite{GR.14}. In the $XY$ chains which will be considered,
i.e., $J_z^{ij}=0$, with $|J_y^{ij}|<J_x^{ij}$ and $J_x^{ij}>0$,  the quantum
discord for the states (\ref{rij}) will always prefer a measurement along the
$x$ axis, irrespective of the field intensity \cite{CRC.10}.

The information deficit (\ref{ID}) can be evaluated in a similar way. In
contrast with the quantum discord, the optimizing measurement direction will be
affected by the field intensity, exhibiting a smooth transition from the $x$ to
the $z$ direction as the field increases for the systems considered,  as
discussed below. The angle $\gamma$ between $\bm{k}$ and the $z$ axis can be
determined from Eq.\ (\ref{stk}), which leads explicitly to
\begin{equation}\cos\gamma=\frac{\alpha_1 r_B+\alpha_2 j_z r_A}{\alpha_3(j_x^2-j_z^2)}\,,
 \label{cg}\end{equation}
when $\gamma\neq 0$ \cite{RCC.11}, which is a transcendental equation (as the $\alpha_i$
depend on $\gamma$).

The quadratic information deficit (\ref{I2}) can, however, be analytically
evaluated with Eq.\ (\ref{I2m}). Here $M_2$ is already diagonal,
${M_2}_{\mu\nu}=\delta_{\mu\nu}(\delta_{\mu z}r_B^2+j_\mu)$. Assuming
$|j_x|\geq |j_y|$, as will occur  in the cases considered, we obtain
\begin{eqnarray} I_2&=&{\textstyle \frac{1}{2}{\rm Min}[j_y^2 +j_x^2,j_y^2+r_B^2 +j_z^2]}\nonumber\\
&=&{\textstyle
4{\rm Min}[\alpha^2+\beta^2,\frac{a_+^2+a_-^2}{4}+\frac{c^2-(a_+-a_-)c+(\alpha-\beta)^2}{2}]}\,,\;\;\; \label{I2a}
\end{eqnarray}
with the minimizing measurement direction $\bm{k}$ along the $z$ ($x$) axis if
the first (second) entry is minimum:
\begin{equation} \bm{k}=\left\{\begin{array}{cc}\bm{e}_z\,,&j_x^2<r_B^2+j_z^2\\
\bm{e}_x\,,&j_x^2>r_B^2+j_z^2\end{array}\right.\,.
\end{equation}
This entails that as the field $B$ increases from $0$, a sharp $x\rightarrow z$
transition in the minimizing measurement direction will take place for $I_2$,
reflecting  the change in the largest eigenvalue of the matrix $M_2$. This
transition becomes softened in the von Neumann information deficit (\ref{ID}),
where $\bm{k}$ will evolve smoothly from the $x$ to the $z$ axis within a
narrow field interval located in the vicinity of the $I_2$ transition. A
measurement transition also occurs for other values of $q$ in the quantities
(\ref{Iq})--(\ref{IRq}) (see \cite{RCC.11} for an example).

\subsection{Results}

\begin{figure}[!ht]
 \centerline{\scalebox{.5}{\includegraphics{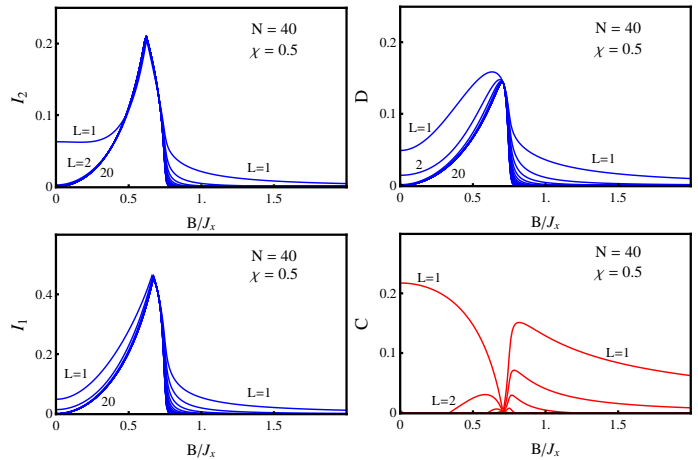}}}
\caption{Left: The one-way information deficits $I_2$ (Eq.\ (\ref{I2}), top)
and $I_1$ (Eq.\ (\ref{ID}), bottom), as a function of the scaled magnetic field
$B/J_x$, for spin pairs with separation $L=1,2,\ldots,n/2$ in the exact ground
state of a cyclic chain of $n=40$ spins with first neighbor anisotropic $XY$
couplings ($\chi=J_y/J_x=1/2$). Right: The quantum discord $D$ (Eq.\ (\ref{D2}), top)
and the concurrence  $C$ (Eq.\ (\ref{Cij}), bottom) for the same pairs. The
results for different separations coincide exactly at the factorization field
$B_s=\sqrt{J_y J_x}\approx 0.71 J_x$.}
 \label{f1}
\end{figure}

\begin{figure}[!ht]
\centerline{\scalebox{.5}{\includegraphics{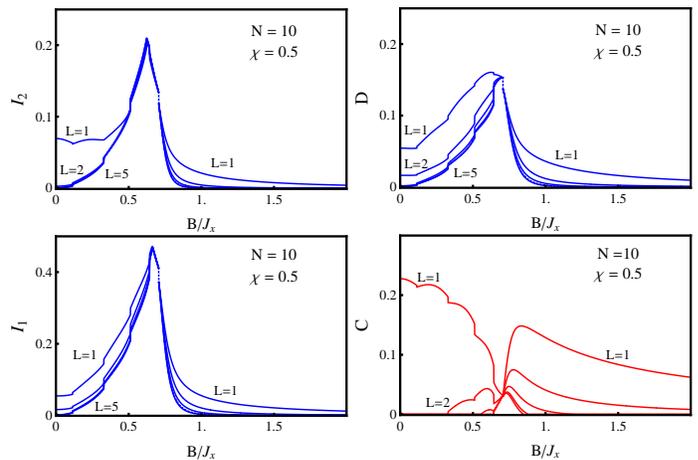}}}
\caption{The same quantities of Fig.\ \ref{f1} for $n=10$ spins. In this case
the parity transitions of the ground state lead to small but appreciable
discontinuities in all quantities, with the last transition (indicated by the
vertical dotted line) taking place at the factorizing field $B_s$. For this
size the concurrence also presents small but finite side limits at $B_s$.}
\label{f2}
\end{figure}

In Figs.\ \ref{f1}--\ref{f2} we show results for the exact  ground state  of a
finite chain with $n$ spins coupled through cyclic ($n+1\equiv 1$) first
neighbor anisotropic $XY$ couplings ($J_z^{ij}=0$, $J_\mu^{ij}=\delta_{j,i\pm
1}J_\mu$ for $\mu=x,y$), for which the reduced pair states (\ref{rij}) will
depend just on the separation $L=|i-j|$ between the spins of the pair. The
exact values of the elements of the density matrix (\ref{rij}) can be obtained,
for any size $n$ or separation $L$, through the Jordan-Wigner fermionization of
the model \cite{LM.61} and its analytic parity dependent diagonalization
\cite{CR.07,RCM.08,PF.09} (see Appendix).

We will set $J_x>0$, with $|J_y|\leq J_x$. This involves no loss of generality
as the sign of $J_x$ can be changed by a local rotation of angle $\pi$ around
the $z$ axis at even sites (assuming $n$ even in cyclic chains), which will not
affect the value of the correlation measures, and the $x$ axis can be chosen
along the direction of maximum coupling.

Fig.\ \ref{f1} depicts the  behavior with increasing field $B$ of the one way
information deficits $I_1\equiv I$ (Eq.\ (\ref{ID}))  and  $I_2$ (Eqs.\
(\ref{I2})-(\ref{I2a})) of spin pairs in the exact definite parity ground state
for the anisotropic case $J_y=J_x/2$, together with that of the quantum discord
(\ref{ID}) and the concurrence (\ref{Cij}). It is first seen that  $I_1$, $I_2$
and $D$  exhibit a similar qualitative  behavior, acquiring appreciable finite
values for {\it any} separation $L$ in the interval $|B|<B_c=(J_x+J_y)/2$, in
marked contrast with the concurrence, which is appreciable just for first and
second neighbors (except for the immediate vicinity of the factorizing field,
see below). The $S_z$ parity symmetry is essential for this result. In fact,
all  measures converge to a {\it finite common value}, independent of the
separation  $L$, at the factorizing field
\cite{Kur.82,RCM.08,GAI.09,RCM.09,CA.13}
\begin{equation}B_s=\sqrt{J_y J_x}\,,\label{Bs}\end{equation}
existing for $0<J_y<J_x$, where the system possesses a pair of degenerate
completely separable exact ground states \cite{RCM.08,RCM.09,CA.13} given by
$|\Theta\rangle=|\theta,\ldots,\theta\rangle$ and
$|-\Theta\rangle=P_z|\Theta\rangle=|-\theta,\ldots,-\theta\rangle$, where
$|\theta\rangle=e^{-i\theta s_y}|\!\downarrow\rangle$ is the single spin state
forming an angle $\theta$ with the $-z$ direction and
$\cos\theta=B_s/J_x=\sqrt{J_y/J_x}$. Actually, in the finite case this field
coincides with  {\it the last parity transition} of the exact (and hence of definite
parity) ground state \cite{RCM.08}, such that the latter approaches, as side limits at
$B=B_s$, the definite parity combinations \cite{RCM.08,RCM.09}
\begin{equation}|\Theta_{\pm}\rangle=\frac{|\Theta\rangle\pm|-\Theta\rangle}
{\sqrt{2(1\pm\langle -\Theta|\Theta\rangle)}}\,.\label{stt}\end{equation}
Here $|\Theta_+\rangle$ ($|\Theta_-\rangle$) is the ground state limit for
$B\rightarrow B_s^+$ ($B\rightarrow B_s^-$). Discarding the overlap $\langle
-\Theta|\Theta\rangle=\cos^n\theta$, which is negligible if $n$  and $\theta$
are not too small ($\cos^n\theta\approx e^{-n\theta^2/2}$ for small $\theta$),
Eq.\ (\ref{stt}) leads to a {\it common} reduced state for {\it any} pair
$i,j$, given by \cite{RCM.08,CRC.10}
\begin{equation}\rho_{\theta}={\textstyle\frac{1}{2}}(|\theta\rangle\langle\theta|
\otimes|\theta\rangle\langle\theta|+
|-\theta\rangle\langle-\theta|\otimes|-\theta\rangle\langle-\theta|)
\,.\label{sttr}\end{equation}
This is a separable mixed state and therefore, it leads to a zero concurrence
for any pair, as seen in Fig.\ \ref{f1} (where results at $B_s$ correspond to
the side limits (\ref{stt})). However, it is not of the classically correlated
form (\ref{4cc}) or (\ref{5cc}) if $\langle-\theta|\theta\rangle=\cos\theta\neq
0$ or $1$, i.e. if $|\pm\theta\rangle$ are non-orthogonal and distinct, leading
then to a common appreciable value of $D$, $I_1$, $I_2$  and in fact all $I_f$.
We also notice that the same reduced state (\ref{sttr}) is obtained from the
mixture $\frac{1}{2}(|\Theta\rangle\langle
\Theta|+|-\Theta\rangle\langle-\Theta|)$, which represents the low temperature
limit of the thermal state $\rho\propto \exp[-H/kT]$ at $B=B_s$.

It is then possible to obtain straightforward analytic expressions for the side
limits of  $D$ \cite{CRC.10}, $I_2$ and  $I_1$ at the factorizing field through
the state (\ref{sttr}),  which leads to $a_{\pm}=\frac{1}{4}(1\pm\cos\theta)^2$
and  $\alpha=\beta=c=\frac{1}{4}\sin^2\theta$ in (\ref{rij}), with
$\cos^2\theta=J_y/J_x$. That for $I_2$ is particularly clean  and given by
\begin{equation}I_2(B_s)=\left\{\begin{array}{cc}\frac{(1-\chi)^2}{2}\,,&\chi\geq 1/3\\
\frac{\chi(1+\chi)}{2}\,,&\chi\leq
 1/3\end{array}\right.\,,\;\;\;\;\chi=J_y/J_x\,,\label{I2bs}\end{equation}
with the minimizing measurement at $B_s$ being along $z$  if $\chi>1/3$ and
along $x$ if $\chi<1/3$. Eq.\ (\ref{I2bs}) applies for all separations $L$.

For small chains, the results are similar but the effects of the parity
transitions of the ground state (it undergoes $n/2$ parity transitions as the
field increases from $0$, the last one at $B=B_s$ \cite{RCM.08}) are now
appreciable trough the finite discontinuities exhibited by $I_2$, $I_1$ and
$D$, as seen in Fig.\ \ref{f2}. At the factorizing field, these discontinuities
arise from the overlap $\langle -\Theta|\Theta\rangle$, which now cannot be
strictly neglected. It leads to an additional term
$\propto\pm\cos^{n-2}\theta(|\theta\rangle\langle-\theta|\otimes|\theta\rangle\langle-\theta|+h.c.)$
in Eq.\ (\ref{sttr}), which originates slightly distinct side limits of $D$
\cite{CRC.10} and also $I_2$ and $I_1$  at $B_s$. Moreover, it also leads to
small but finite and distinct common side limits of the concurrence at $B=B_s$
\cite{RCM.08,RCM.09}, which was known to reach full range in its vicinity
\cite{Am.06}.  All these side limits are, nevertheless, still
 independent of the pair separation $L$. In the case of $I_2$, they are given,
for $\chi\gtrsim 1/3$, by
\begin{equation}
I_2(B_s^{\pm})=
\frac{(1-\chi)^2}{2}\frac{1+\chi^{n-2}}{(1\pm\chi^{n/2})^2}\,,\label{I2bsn}\end{equation}
which corrects the upper line in Eq.\ (\ref{I2bs}) for finite $n$ (or
$\chi\rightarrow 1$) and $+$ ($-$) corresponds to the right (left) side limit.
The side limits of the concurrence are
$C(B^{\pm}_s)=\frac{\chi^{n/2-1}(1-\chi)}{1\pm\chi^{n/2}}$, as obtained from
(\ref{Cij}) \cite{RCM.08,RCM.09}.

The behavior of the quantum discord  for longer range ferromagnetic-type
couplings is qualitatively similar \cite{CRC.10}. Moreover, a factorizing field
still exists for longer range couplings with a constant  anisotropy
$\chi=J_y^{ij}/J_x^{ij}$ \cite{RCM.09}, in which case the reduced pair state
at $B_s$ is again given by Eq.\ (\ref{sttr}) with $\cos\theta=\sqrt{\chi}$, and
Eqs.\ (\ref{I2bs})--(\ref{I2bsn}) remain then valid.

\begin{figure}[t]
\centerline{\scalebox{.5}{\includegraphics{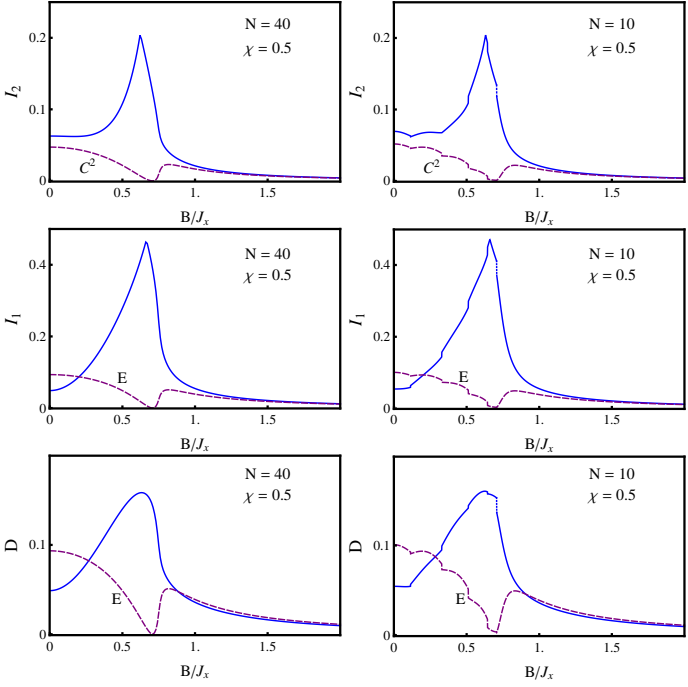}}} \vspace*{0.cm}
\caption{Plot of $I_2$ (top), $I_1$ (center), and $D$
(bottom) together with the associated entanglement monotones for a first neighbor pair
($L=1$) in the ground state of the chains of Figs.\ \ref{f1} and \ref{f2}.}
\label{f3}
\end{figure}

In  Fig.\ \ref{f3} we compare the behavior of  $I_2$, $I_1$ and $D$ for first
neighbors in the chains of Figs.\ \ref{f1} and \ref{f2}, with that of the
associated entanglement monotone, i.e., the squared concurrence $C^2$ for $I_2$
and the  entanglement of  formation $E$ for $I_1$ and $D$, such that both
quantities coincide for pure states. It is seen that for strong fields,
differences are very small, in agreement with the weak entanglement of the pair
with the rest of the chain in this regime  ($\rho_{i,i+1}$ is almost pure). The
strong differences arise for $B<B_c$, and especially in the vicinity of the
factorizing field, due to the arguments exposed above. For $|B|<B_c$ the reduced
pair state becomes appreciably mixed in the definite parity ground states,
including the states (\ref{stt}) at the factorizing field, due to the
entanglement with the rest of the chain. Significant differences between $I_f$
(and $D$) with the corresponding entanglement monotone become then feasible.

It is also seen that $I_2$ is in this case an upper bound of  $C^2$ for all
fields, whereas  $I_1$  is not a upper bound of $E$ for low fields while  $D$ is
not a upper bound even for strong fields, indicating the lack of an order
relationship between $D$ and $E$ even in this regime. In the case of $I_2$, it
is easy to show from Eqs.\ (\ref{Cij}) and (\ref{I2a}) that for $X$ states, it
is always an upper bound of $C^2$ when the minimizing measurement is along $z$ \cite{RCC.11}.
In, fact, for strong fields $|B|\gg J_x$, a perturbative expansion
\cite{RCC.10} for the present chain  leads to $C\approx 2(\eta-\eta^2)$,
$I_2\approx 4\eta^2$, $I_1\approx \eta^2(\log e -\log\eta^2)$ and $D\approx
\eta^2(\log e-\log\eta^2-2)$, where
\[\eta=\frac{J_x-J_y}{8 B}\,.\]
Hence, in this limit $I_2-C^2=O(\eta^3)$ and $I_1-E=O(-\eta^3\log \eta^2)$, both positive,
whereas $D-E\approx O(-\eta^2)$ becomes negative.

\subsection{Minimizing measurement}

\begin{figure}[ht!]
 \centerline{\scalebox{.9}{\includegraphics{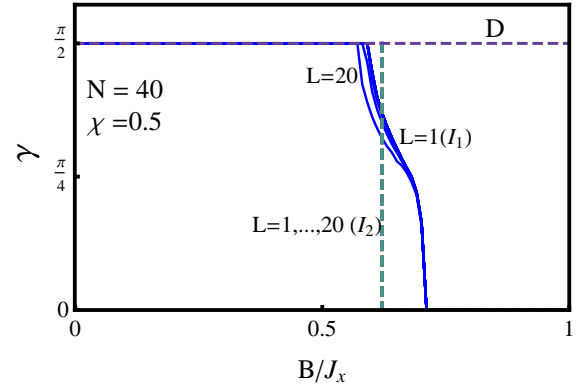}}}
\caption{The angle $\gamma$  determining the direction of the minimizing local
spin measurement for $D$,  $I_1$ and  $I_2$, as a function of the scaled
transverse magnetic field, for a chain of  $n=40$ spins with $J_y=J_x/2$.
Results for all separations $L$ of the pair are shown.} \label{f4}
\end{figure}

Although $I_1$, $I_2$ and $D$ show a similar qualitative  behavior, both
measures $I_1$, and  $I_2$ exhibit a more pronounced maximum, in comparison to
that of  the quantum discord, as appreciated in  Figs.\ \ref{f1}-\ref{f3}. This
reflects the transition in the orientation of their local minimizing spin
measurements as the field increases, which, as mentioned above, is not present
in the quantum discord. The latter prefers in the  present system a measurement
along the $x$ axis, even for large fields and for any separation between the
spins, following the strongest correlation \cite{GR.14}. As seen in Fig.\
\ref{f4} and as previously stated, $I_2$  exhibits instead a sharp transition
from a direction {\it parallel to the $x$ axis} ($\gamma=\pi/2$) to a direction
{\it parallel to the  $z$ axis} ($\gamma=0$) i.e., parallel to the field.  This
transition  takes place, in the case shown in  Fig.\ \ref{f1}, for all
separations $L$ at $B\approx 0.65 J_x$. In the case of the Information Deficit
$I_1$, the transition  becomes smooth,  as the angle $\gamma$ takes all the
intermediate values between $ 0$ and $\pi/2$ (as determined by Eq.\ (\ref{cg}))
for all separations in a narrow field interval centered at the $I_2$ critical
field, as also seen in Fig.\ \ref{f4}.

\begin{figure}[ht!]
\centerline{\scalebox{.8}{\includegraphics{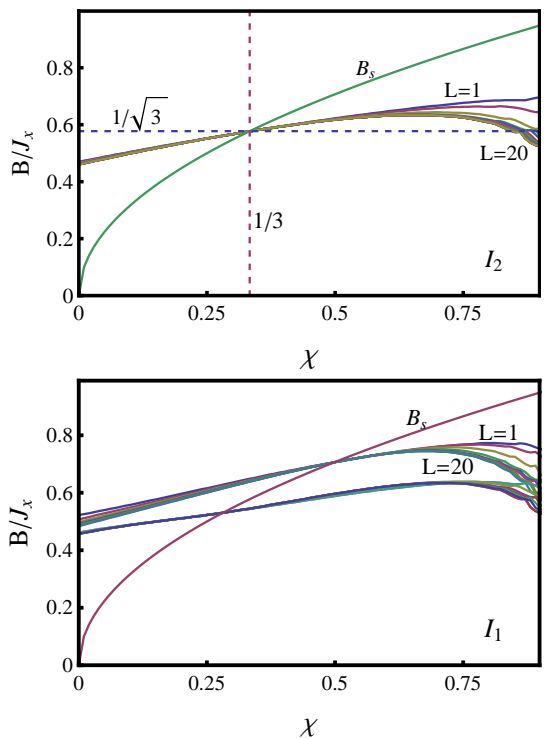}}}
 \vspace*{0.1cm}
\caption{Top: The field where the  transition in the minimizing measurement of
$I_2$ takes place, as a function of the  anisotropy $\chi=J_y/J_x$. A direction
along the $x$ ($z$) axis is preferred below (above) the transition field. The
factorizing field  $B_s$ is also shown. All transition fields coincide with the
factorizing field if $\chi=1/3$ (Eq.\ (\ref{I2bs})). Bottom: The fields
delimiting the interval where the smoothed transition in the minimizing
measurement of the von Neumann information deficit $I_1$ takes place. }  \label{f5}
\end{figure}

The value of the field where  the transition in the optimizing local
measurement for $I_2$ occurs, depends on the anisotropy but only slightly on
the separation $L$, except in the $XX$ limit ($J_y\rightarrow J_x$),  as can be
seen in the top panel of Fig.\ \ref{f5}. The same holds for the field interval
where the ``transition'' (actually the evolution from $\pi/2$ to $0$ of the
measurement angle $\gamma$) in $I_1$ takes place (bottom panel of Fig.\
\ref{f5}). In the case of $I_2$, if $\chi=1/3$ the measurement transition for
{\it all} separations $L$ occurs {\it exactly} at the factorizing field
$B_s=\sqrt{\chi}J_x$, as follows from Eq.\ (\ref{I2bs}).

The measurement transition reflects essentially the qualitative change experienced by
the reduced state of the pair for increasing fields. Away from the $XX$ limit, the
dominant eigenstate of $\rho_{ij}$ (that with the largest eigenvalue) for not
too low fields is the entangled state
$|\Psi_+\rangle=u|\!\!\downarrow\downarrow\rangle+v|\!\!\uparrow\uparrow\rangle$
with $v/u=\frac{\beta}{\varepsilon +\sqrt{\varepsilon^2+\beta^2}}$ and
$\varepsilon=\frac{a_--a_+}{2}$. Above the measurement transition field (i.e.,
when the optimum measurement is parallel to the field),  $v/u$ becomes
small ($\lesssim 0.25$), indicating that the pair is approximately aligned with
the field. Instead, below the transition field $v/u$ increases,
approaching $1$ for $B\rightarrow 0$ (where $|\Psi_+\rangle$ becomes a
parallel Bell state) and the least disturbing measurement is along $x$. For
very low fields the dominant eigenstate may shift to the antiparallel Bell
state $|\Psi_-\rangle=\frac{|\uparrow\downarrow\rangle+|\downarrow\uparrow\rangle}{\sqrt{2}}$
arising from the central block of (\ref{rij}), and in this case the measurement
along $x$ is still preferred. On the other hand, in the $XX$ limit, $\beta=0$
in (\ref{rij}) and  the dominant eigenstate is either $|\Psi_-\rangle$
at low fields, or $|\Psi_+\rangle=|\!\downarrow\downarrow\rangle$ for strong fields,
and the measurement transition of $I_2$ indicates essentially the field where
the sharp transition in the dominant eigenstate (from maximally entangled to
separable) takes place \cite{CCR.13}. Such measurement transition for increasing fields
 persists even at finite temperatures \cite{CCR.13}.

\section{Conclusions}

We have examined the behavior of the quantum discord and the standard and
quadratic one-way information deficit of spin pairs in the exact definite
parity ground state of a finite anisotropic cyclic $XY$ spin $1/2$ chain in a
transverse field. We have first provided a brief overview of the quantum
discord, the standard von Neumann based one-way information deficit and the
generalized information deficit, which contains the standard as well the
quadratic deficit as particular cases, and which can be interpreted as a
measure of the minimum entanglement generated between the system and the
measurement apparatus after a complete local projective measurement.  The first
important result is that the behavior of all these measures is quite distinct
from that of the pair  entanglement  for fields below the critical field,
acquiring  finite appreciable values for {\it all separations} of the spins of
the pair. Moreover, they  reach (as side limits) a common (independent of the
separation) finite value at the factorizing field, which in a finite chain is
the field where the last ground state parity transition takes place. These
finite limits can be evaluated analytically. The entanglement of pairs also
reaches full range in its vicinity, although its  value is much smaller and
vanishes at this field except for very small samples. Parity effects are of
crucial importance  for the proper description of these measures in finite
systems below the critical field.

The second important result is that the behavior of the optimizing local spin
measurement of both the standard and generalized information deficit is quite
distinct from that optimizing the quantum discord, exhibiting a transition
in the direction of the spin measurement, from that of maximum correlation
to that parallel to the field. The details of this transition depend on
the choice of entropy (it is sharp for $I_2$, and smooth for $I_1$).  The
quantum discord prefers instead that of maximum correlation  even for strong
fields. Hence, the quantum discord, which is based on the minimization of a
conditional entropy, ``detects'' in this way this direction \cite{GR.14},
while the information deficits, based on the minimization of a total entropy,
are more sensible to changes in the structure of the reduced state of the pair.

A final comment is that the generalized formalism permits the use of simple
entropic forms involving just low powers of the density matrix, leading to
measures of the form (\ref{Iq}) or (\ref{IRq}) which can be more easily
evaluated and optimized, and which are also more easily accessible from the
experimental side.

\appendix
\section{Appendix}

We briefly discuss here the exact solution of the {\it finite} cyclic $XY$
chain with first neighbor couplings, which requires to take  into account
exactly the parity effects \cite{LM.61,CR.07,RCM.08,PF.09}. The Jordan Wigner
transformation \cite{LM.61} allows to rewrite the Hamiltonian (\ref{H}) in the
$XY$ case ($J_z^{ij}=0$) for $J_\mu^{ij}=J_\mu\delta_{i,j\pm 1}$, $\mu=x,y$,
and for each value $\pm 1$ of the $S_z$ parity $P_z$, as a quadratic form in
fermion creation and annihilation operators $c^\dagger_i$, $c_i$ defined by
$c^\dagger_i=s_{i+}\exp[-i\pi\sum_{j=1}^{i-1}s_{j+}s_{j-}]$, with the reverse
transformation given by
$s_{i+}=c^\dagger_{i}\exp[i\pi\sum_{j=1}^{i-1}c^\dagger_{j}c_{j}]$. This leads
to
\begin{eqnarray} H^{\pm}&=&
\sum_{i=1}^n B(c^\dagger_ic_i-{\textstyle\frac{1}{2}})-
{\textstyle\frac{1}{2}}\eta^{\pm}_i(J_+c^\dagger_i c_{i+1}
+J_-c^\dagger_i c^\dagger_{i+1}+h.c.)\nonumber\\&=&\sum_{k\in K_{\pm}}\!\!\lambda_k (a^\dagger_k a_k
 -{\textstyle\frac{1}{2}})\label{qd}\end{eqnarray}
where  $J_{\pm}={\textstyle\frac{1}{2}}(J_x\pm J_y)$ and $n+1\equiv 1$,
$\eta^-_i=1$, $\eta^+_i=1-2\delta_{in}$ \cite{LM.61}. In (\ref{qd}),
$K_+=\{{\textstyle\frac{1}{2}},\ldots,n-{\textstyle\frac{1}{2}}\}$,
$K_-=\{0,\ldots,n-1\}$ and
 \begin{equation}\lambda_k=\sqrt{(B-J_+\cos\omega_k)^2+J_-^2\sin^2\omega_k}\,, \;\;\;
 \omega_k=2\pi k/n\,. \end{equation}
The last form (\ref{qd}) is obtained through a parity dependent discrete
Fourier transform $c^{\dagger}_j=\frac{e^{i\pi/4}}{\sqrt{n}} \sum\limits_{k\in
K_{\pm}} e^{-i\omega_k j}c'^\dagger_k$, followed by a BCS-type Bogoliubov transformation
$c'^\dagger_k=u_k a^\dagger_k+v_ka_{n-k}$, $c'_{n-k}=u_k a_{n-k}-v_ka^\dagger_k$ to
quasiparticle fermionic operators  $a_k$, $a^\dagger_k$, with
$(^{u_k^2}_{v_k^2})={\textstyle\frac{1}{2}}[1\pm(B-J_+\cos\omega_k)/\lambda_k]$.

For $B\geq 0$, we may set $\lambda_k\geq 0$ for $k\neq 0$ and
$\lambda_0=J_+-B$, in which case the quasiparticle vacuum of $H^{\pm}$ has the
right parity and the lowest energy is
$E^{\pm}=-{\textstyle\frac{1}{2}}\sum_{k\in K_{\pm}}\lambda_k$. At the
factorizing field (\ref{Bs}), $\lambda_k=J_+-B_s\cos\omega_k$ and
$E^\pm=-nJ_+/2$ \cite{RCM.08}.

The reduced state of a spin pair in the exact ground state can then be
obtained from the basic contractions $\langle a^\dagger_k a_{k'}\rangle=0$,
$\langle a^\dagger_{k}a^\dagger_{k'}\rangle=0$, leading to $\langle {c'}^\dagger_k
{c'}_{k'}\rangle=v_k^2\delta_{kk'}$, $\langle {c'}^\dagger_k
{c'}^\dagger_{k'}\rangle=u_kv_k\delta_{k,-k'}$ and  ($L=i-j$)
\begin{eqnarray}\langle c^\dagger_ic_j\rangle_{\pm}&=&\frac{1}{n}\sum_{k\in K_{\pm}} e^{-i\omega_k  L}
v_{k}^2=f_L+{\textstyle\frac{1}{2}}\delta_{ij}\,,\nonumber\\
\langle c^\dagger_i c^\dagger_j\rangle_{\pm}
&=&\frac{1}{n}\sum_{k\in K_{\pm}} e^{-i\omega_k L}u_k    v_k=g_L\,.\end{eqnarray}
Application of Wick's theorem then leads to  \cite{LM.61,CRC.10}
\begin{eqnarray}\langle s_{iz}\rangle&=&f_0,\;\langle
s_{iz}s_{jz}\rangle=f_0^2-f_L^2+g_L^2\,,\nonumber\\
\langle s_{i-}s_{j\mp}\rangle&=& {\textstyle\frac{1}{4}[{\rm det}(A^+_L)
 \mp{\rm  det}(A^-_L)]} \,,\nonumber\end{eqnarray}
where  $(A_L^\pm)$ are $L\times L$ matrices of elements
$(A_L^{\pm})_{ij}=2(f_{i-j\pm 1}+g_{i-j\pm 1})$. These results, valid for any
finite $n$, were checked through direct diagonalization for small $n$.

\acknowledgments{
The authors acknowledge support form CONICET  (L.C. and N.C.) and CIC (R.R.)
of Argentina.}

\end{document}